\documentclass[pre,twocolumn,showpacs,superscriptaddress] {revtex4}
\usepackage{graphics}
\newcommand{\sinp}
{\affiliation{Theoretical Condensed Matter Physics Division and
Center for Applied Mathematics and Computational Science,\\
 Saha Institute of Nuclear Physics,
 1/AF Bidhannagar, Kolkata 700064, India}}
\newcommand{\cu}
{\affiliation{Department of Physics, University of Calcutta,
92 Acharya Prafulla Chandra Road, Kolkata 700009, India}}

\begin{document}
  
 \title
 {
Zero temperature  dynamics in two dimensional ANNNI model}
 \author
 {
Soham Biswas}
\cu
\author { Anjan Kumar Chandra}
\sinp
\author	{  Parongama Sen }
\cu
\begin{abstract}
We investigate  the  dynamics of a two dimensional axial next nearest neighbour
Ising  (ANNNI) model following
a quench to zero temperature. The Hamiltonian is given by
$H = -J_0\sum_{i,j=1}^L S_{i,j}S_{i+1,j} - J_1\sum_{i,j=1} [S_{i,j} S_{i,j+1} -\kappa S_{i,j} S_{i,j+2}]$.
For $\kappa <1$, the system does not reach the 
equilibrium ground state but slowly evolves to a metastable state.
For $\kappa > 1$, the system shows a behaviour similar to the 
two dimensional ferromagnetic Ising model in the sense that it freezes to 
a striped state with a finite probability. The persistence probability shows algebraic 
 decay here with an exponent $\theta = 0.235 \pm 0.001$ while the
dynamical exponent of growth $z=2.08\pm 0.01$.
For $\kappa =1$, the system belongs to a completely different
dynamical class; it always evolves to the true ground state with  the
persistence and dynamical exponent having unique values.
Much of the dynamical phenomena can be
understood by  studying the dynamics and distribution of the number of domains walls. 
We also compare the dynamical behaviour to that of a Ising   model in which both the nearest and 
next nearest neighbour interactions are ferromagnetic.

\vskip 0.5cm

\end{abstract}

\maketitle
\section{Introduction}

 Dynamics of Ising models  is a much studied
phenomenon and has emerged as a rich field of
present-day research. Models having identical static critical behavior may display different
behavior when dynamic critical phenomena are considered \cite{Ho_Ha}.
An important dynamical feature commonly studied is the quenching
phenomenon below the critical temperature.
In a quenching process, the system has a disordered initial configuration
corresponding to a high temperature and its temperature is suddenly dropped.
This results in quite a few interesting phenomena like domain growth \cite{gunton,bray},
persistence \cite{satya1,derrida,stauffer,krap1,Krap_Redner} etc.

In  one dimension, a zero temperature quench  of the  Ising model
ultimately leads to the equilibrium
configuration, i.e., all spins point up (or down).
 The average domain size $D$ increases in time $t$ as $D(t)\sim t^{1/z}$,
where $z$ is the dynamical exponent associated with the growth.
As the system coarsens, the magnetisation also
grows in time as $m(t)\sim t^{1/2z}$.
In two or higher dimensions, however, the system does not
always reach equilibrium \cite{Krap_Redner} although these scaling relations
still hold good.

  Apart from the domain growth  phenomenon, another important dynamical
  behavior commonly studied is persistence. In Ising model, in a zero temperature quench, persistence 
is simply the probability that a spin has not flipped till time $t$ and
is given
by $P(t)\sim t^{-\theta}$. $\theta$ is called the persistence exponent and
is unrelated to any other known static or dynamic exponents. 

Drastic changes in the dynamical behaviour of the Ising model
in presence of a competing next nearest neighbor interaction have been observed 
earlier \cite{redner,sdg_ps,barma}. 
The one dimensional ANNNI (Axial next nearest neighbour Ising) model with $L$ spins is described by the Hamiltonian
\begin{equation}
H = -J\sum_{i=1}^L (S_iS_{i+1} - \kappa S_i S_{i+2}).
\end{equation}
Here it  was found
that for $\kappa < 1$,  under a zero temperature quench with
single spin flip Glauber dynamics,   the system does not reach  its
true ground state. (The ground state is ferromagnetic for
$\kappa < 0.5$, antiphase for $\kappa> 0.5$, and highly degenerate at $\kappa=0.5$ \cite{selke}). On the contrary, after an initial short time,
domain walls become fixed in number but remain mobile 
at all times thereby making the persistence probability go to
zero in a stretched exponential manner. For $\kappa > 1$ on the other hand, 
although the
system reaches the ground state at long times, the dynamical
exponent and the persistence exponent are both different from 
those of the Ising model with only nearest neighbour interaction \cite{sdg_ps}.

The above observations and  the additional fact that even in the two dimensional nearest neighbour 
Ising model, 
frozen-in striped states appear in a zero temperature quench \cite{Krap_Redner}, 
suggest that the   two 
dimensional Ising model 
in presence 
of competing interactions could show novel dynamical behaviour. In the present work, we have introduced such
an interaction (along one direction) 
in the two dimensional Ising model, thus making it equivalent 
to the ANNNI model 
in two dimensions
precisely.  The Hamiltonian for the two dimensional ANNNI model 
on a  $L\times L$ lattice is given by
\begin{equation}
H = -J_0\sum_{i,j=1}^L S_{i,j}S_{i+1,j} - J_1\sum_{i,j=1} [S_{i,j} S_{i,j+1} -\kappa S_{i,j} S_{i,j+2}].
\label{annni2d}
\end{equation} 
Henceforth, we will assume the competing interaction to be
along the $x$ (horizontal) direction, while in the $y$ (vertical) direction,
there is only ferromagnetic interaction. 

\begin{figure}[h]
{\resizebox*{6cm}{!}{\includegraphics{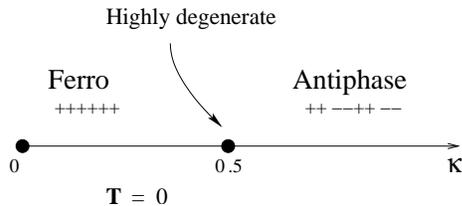}}}
\caption{The ground state (temperature $T=0$) 
spin configurations along the $x$ direction are shown 
for different values of  $\kappa$. In the ferromagnetic phase, there is a two fold
degeneracy and in the antiphase the degeneracy is four fold. The ground state is infinitely degenerate at the fully frustrated point $\kappa = 0.5$.}  
\label{phase}
\end{figure}

Although the thermal phase diagram of the two dimensional ANNNI model
is not known exactly,  the ground state is known and simple. 
If one calculates the
magnetisation along the horizontal direction only, then 
for $\kappa< 0.5$, there is ferromagnetic order and antiphase order for $\kappa > 0.5$. Again, $\kappa=0.5$ is the fully
frustrated point where the ground state is 
highly degenerate. On the other hand, there is always ferromagnetic order along the
vertical direction.
In Fig. \ref{phase}, we have shown the   ground state spin configurations  along the 
$x$ direction for different values of $\kappa$.

In section II, we have given a list of the quantities calculated. 
In section III, we discuss the dynamic behaviour in detail.
In order to compare the results with those of a model without competition,
we have also studied the dynamical features of a two dimensional
Ising model with ferromagnetic next nearest  neighbour interaction, i.e., 
the model given by eq. (\ref{annni2d}) in which $\kappa < 0$. 
These results are also presented in section III.
Discussions and concluding statements are made in the
last section.

\section{QUANTITIES CALCULATED}

We have estimated the following quantities in the present work:

\begin{enumerate}
\item Persistence probability $P(t)$: As already mentioned, 
this is the probability that a spin
 does not flip till time $t$. 

In case the persistence probability shows a power law form,
$P(t) \sim t^{-\theta}$,
one can use the finite 
size scaling relation \cite{puru}
\begin{equation}
P(t,L) \sim t^{- \theta}f(L/t^{1/z}).
\label{fss}
\end{equation}
For finite systems, the 
persistence probability saturates at a value $L^{-\alpha}$ at large times. 
Therefore,  for
 $x <<1$ , $f(x)  \sim x^{-\alpha}$ with $\alpha = z\theta$.  For large $x$,
$f(x)$ is a constant.

\begin{figure}[h]
{\resizebox*{6cm}{!}{\includegraphics{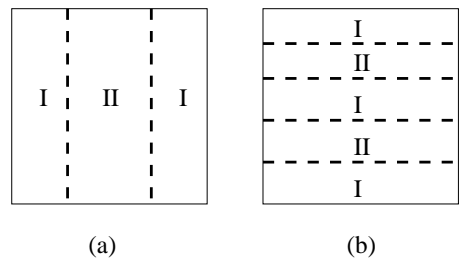}}}
\caption{The schematic pictures of  configurations with flat interfaces
 separating domains of type I and II  are shown:
 (a) when the interface lies parallel to $y$ axis, we have
nonzero $f_{D_x}$ ($=2/L$ in this particular case) and (b) with
interfaces parallel to the  $x$ axis we have nonzero $f_{D_y}$ ($=4/L$ here)}
\label{dxyschematic}
\end{figure}

It has been shown that the exponent $\alpha $ is related to the fractal dimension of the
fractal formed by the persistent spins
 \cite{puru}. Here we obtain an estimate 
of $\alpha$ using the above analysis.
%

\item Number of domain walls  $N_D$: Taking a single strip of $L$ spins at a time, one can calculate the number of domain walls for each strip and determine the average.
In the $L \times L$
lattice, we consider the fraction $f_D = N_D/L$ and 
study the  behaviour of $f_D$ as a function of time.
One can  take strips  along both the $x$ and $y$ directions (see Fig. \ref{dxyschematic}
 where the
calculation of $f_D$ in simple cases has been illustrated). As the system
is anisotropic, it is expected that the two measures, $f_{D_x}$ along the $x$ 
direction and $f_{D_{y}}$ along the $y$ direction,  will show different 
dynamical
behaviour in general.   
The domain size $D$  increases as $t^{1/z}$ as already mentioned and it has been observed earlier that the dynamic exponent occurring 
in 
coarsening dynamics is the same as that occurring in the finite size scaling
of $P(t)$  (eq. (\ref{fss})) \cite{puru}. 
Although we do not calculate the 
domain sizes, the average number of domain walls per strip
is shown to follow a dynamics given by the same exponent $z$, at least
for $\kappa > 1$.

\item Distribution $P(f_D)$ (or $P(N_D)$) of the fraction (or number) 
of domain walls 
at steady state:  this  is also done for both $x$ and $y$ directions.

\item Distribution $P(m)$ of the total magnetisation 
at steady state for $\kappa\leq 0 $ only. 

 We have taken  lattices of size $ L\times L$ with $L=40$,$~100$,$~200$ and $300$ 
to study the persistence behaviour and dynamics of the domain walls of the system and averaging over at least 50 configurations for each size have been made.
For estimating the distribution  $N_{D}$ we have averaged over much larger 
number of configurations (typically 4000) and restricted to system sizes 
 $40\times40$, $60\times60$, $80\times80$ and $100\times100$.
Periodic boundary condition has been used in both $x$ and $y$ directions.
 $ J_{0}= J_{1}= 1$ has been used in the numerical simulations.

\end{enumerate}
\section{DETAILED DYNAMICAL BEHAVIOUR} 

Before going in to the details of the dynamical behaviour let us
discuss the stability of simple configurations or structures of spins 
 which will help
us in appreciating the fact that the dynamical behaviour is strongly
dependent on $\kappa$.\\

\subsection{Stability of simple structures}

An important question that arises in dynamics is the stability of 
spin configurations - it may happen that configurations which do not correspond to global
minimum of energy still remain stable dynamically.
This has been termed  ``dynamic frustration'' \cite{pratap} earlier.
A known example is of course a striped state occurring in the
two or higher dimensional Ising models which is stable but not a configuration 
which has minimum energy.

In ANNNI model, the stability of the configurations depend 
very much on the value of $\kappa$. It has been previously analysed
for the one dimensional ANNNI model that $\kappa=1$ is a special point above and below which the dynamical behaviour changes completely
because of the stability of certain  structures in the system.\\

Let us consider the simple configuration of a single up spin
in a sea of down spins. Obviously,  it will be unstable as 
long as $\kappa < 2$. For $\kappa > 2$, although this spin is
stable, all the neighbouring spins are unstable.
However, for $\kappa < 2$, only the up spin is unstable 
and the dynamics will stop once it flips. 
When $\kappa =2$ the spin may or may not flip,
 i.e.,  the dynamics is stochastic.

Next we consider a domain of two up spins in a sea of down spin.
These two may be oriented either along horizontal or vertical direction.
These spins will be stable for $ \kappa > 1$
 only 
while all the neighouring spins are  unstable.
For $\kappa < 1$, all spins except the up spins are stable. When $\kappa = 1$,
 the dynamics is again stochastic.

\begin{figure}[h]
{\resizebox*{6cm}{!}{\includegraphics{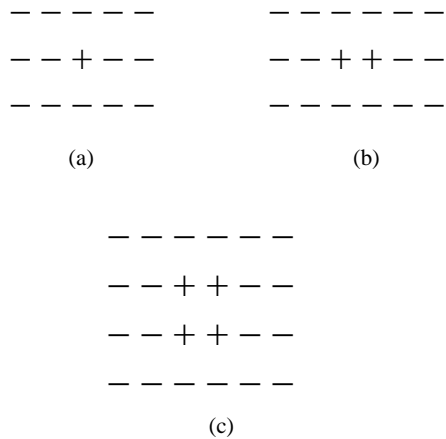}}}
\caption{Analysis of stability of simple structures: (a) single up spin
in sea of down spins; here for $\kappa < 2$ all the spins except the
up spin is stable (b) two up spins in a sea of down spins, all spins except the two
up spins are stable for $\kappa < 1$ (c) a two by two structure of up spins - here
all the spins are stable for $\kappa < 1$ while neighbouring spins are not (see text for details).}
\label{stability}
\end{figure}

A two by two structure of up spins in a sea of down spins on the other
hand will be stable for any value of $\kappa > 0$. But the neighbouring spins along the vertical 
direction will be unstable for $\kappa \geq 1$. This
shows that for $\kappa < 1$, one can expect that the 
dynamics will affect the minimum number of spin and therefore
the dynamics will be slowest here. A picture of the structures described above
are shown in Fig \ref{stability}.

One can take more complicated structures but the analysis of these simple ones
is sufficient  to expect that there will be different dynamical
behaviour in the regions $\kappa < 1, \kappa = 1, \kappa > 1,
\kappa =2$ and  $\kappa > 2$. However, we find that 
as far as persistence behaviour is concerned, there are only three regions
with different behaviour:
$\kappa < 1, \kappa = 1$ and $ \kappa > 1$. On the other hand, 
 when the distribution 
of the number  of domain walls in the steady state is considered,
 the three regions 
 $ 1< \kappa < 2, \kappa = 2$ and $ \kappa > 2$ have clearly distinct behaviour.

\subsection{$ 0 < \kappa < 1$}

\begin{figure}[h]
\centering
\resizebox*{6cm}{!}{\includegraphics{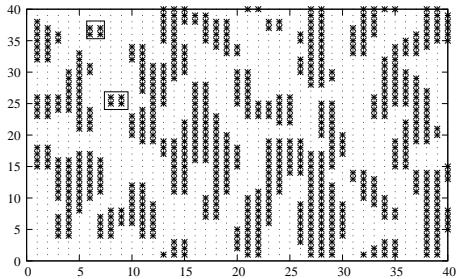}}
 \caption{
Snap shot of a $40 \times 40$ system at time $t= 10$ for $\kappa<1$
A few  simplest quasi frozen structures are highlighted.}
\label{snap1}
\end{figure}

\begin{figure}[h]
\centering
\resizebox*{6cm}{!}{\includegraphics{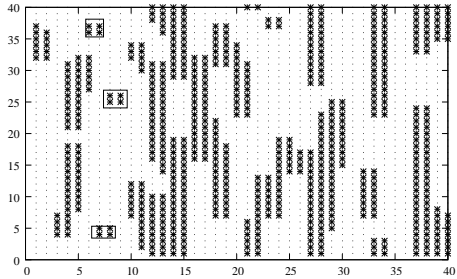}}
\caption{
Same as  Fig. \ref{snap1}  with  $t= 100$.
}
\label{snap2}
\end{figure}

\begin{figure}[h]
\centering
\resizebox*{6cm}{!}{\includegraphics{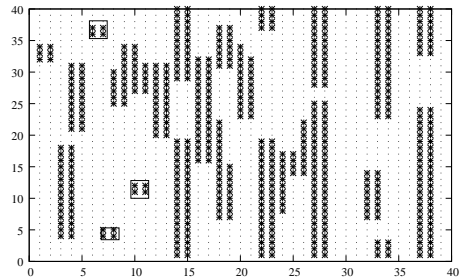}}
\caption{
Same as Fig. \ref{snap1}  with  $t= 500$.
 One of the two by two structures has melted while
another one has formed.
The ladder like structures which have formed are perfectly stable.
}
\label{snap3}
\end{figure}

\begin{figure}[h]
\centering
\resizebox*{6cm}{!}{\includegraphics{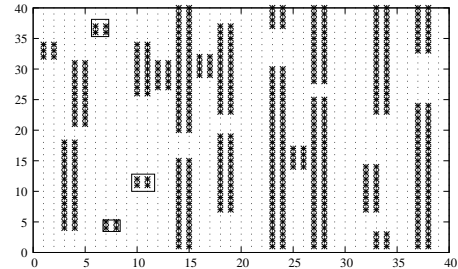}}
\caption{
Same as Fig.  \ref{snap1}  with  $t= 75000$.
This snapshot is taken after a very  long time to show that the 
system has undergone nominal changes compared to the  length of the time interval. The 
whole configuration now consists of ladders and the dynamics 
stops once the system reaches such a state.
}
\label{snap4}
\end{figure}

We find that as in \cite{sdg_ps}, in the region $0 < \kappa < 1$, the system has identical 
dynamical behaviour for all $\kappa$. Also, like the one dimensional
case, here  the
system does not go to its equilibrium ground state. However, the dynamics
continues for a long time, albeit very slowly for reasons mentioned above. 
In Figs. \ref{snap1} - \ref{snap4}, we show the snapshots of the system at different
times for a typical quench to zero temperature.
As already mentioned, here domains of size one and two will vanish very fast
and certain
structures,  the  smallest of which is a two by two domain 
of up/down
spins in a sea of oppositely oriented spins can survive till 
very long times. These structures we call quasi-frozen as the spins inside 
these structures  (together with
the neighbourhood spins)   are locally
stable;  they can be disturbed only  when the effect of 
a spin flip occurring at a distance propagates to its vicinity which 
usually takes a long time.

The pictures at the later stages also
show that the  system tends to attain a configuration in which the domains have
straight  vertical edges, it can be easily checked that structures with kinks are not stable.
 We find a tendency to form strips of width two (``ladders'')
along the
vertical direction - this is due to the second neighbour interaction -
however,
these strips  do not span the entire lattice in general.
The domain structure
is  obviously not symmetric, e.g., ladders along the horizontal direction
will not form stable structures.
The dynamics stops once the entire lattice is spanned by only
ladders of height ${\cal{N}} \leq L$.

\vskip 1.5cm

The persistence probability for $\kappa <1$ shows a very slow decay 
with time which can be approximated by $\frac {1}{\log(t)}$ for 
an appreciable range of time. At later times, it approaches
a saturation value in an even slower manner.
The slow dynamics of the system accounts for this slow decay.

\begin{figure}[h]
\centering
\rotatebox{270}{\resizebox*{6cm}{!}{\includegraphics{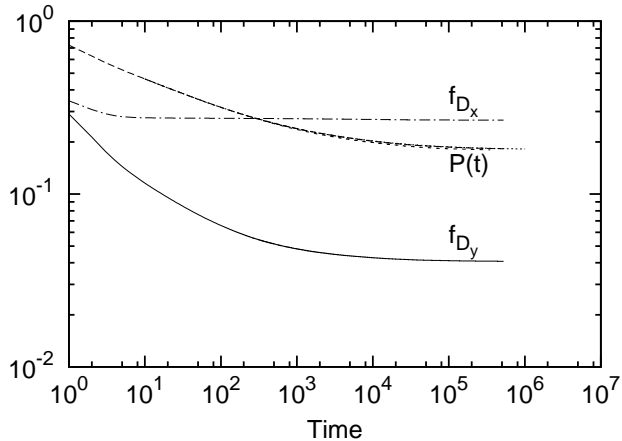}}}
\caption{
Persistence $P(t)$ and average number of domain walls per site, $f_D$ are shown for  $\kappa<1$.
}
\label{dynklt1}
\end{figure}

 The fraction of   domain walls $f_{D_x}$ and $f_{D_y}$ 
along the $x$ direction and $y$ directions
show remarkable difference as  functions of time. While that in the $x$ direction saturates quite fast, in the $y$ direction, 
it shows a gradual 
decay till very long times (see Fig. \ref{dynklt1}). This indicates 
that the dynamics essentially keeps the number of domains unchanged  along
$x$ direction while that in the other direction changes slowly
in time. The behaviour of $f_{D_x}$  is similar to what 
happens in one dimension. In fact,
the average number of domain walls $N_{D_x}$ at large times is also 
very close to that obtained for the ANNNI chain, it is about $0.27L$. 
However, in contrast to the one dimensional case where the domain walls 
remain mobile, here the mobility of the domain walls are impeded  by
the presence of the ferromagnetic interaction along the vertical direction
causing a kind of pinning of the domain walls.

\begin{figure}[h]
\centering
\rotatebox{270}{\resizebox*{6cm}{!}{\includegraphics{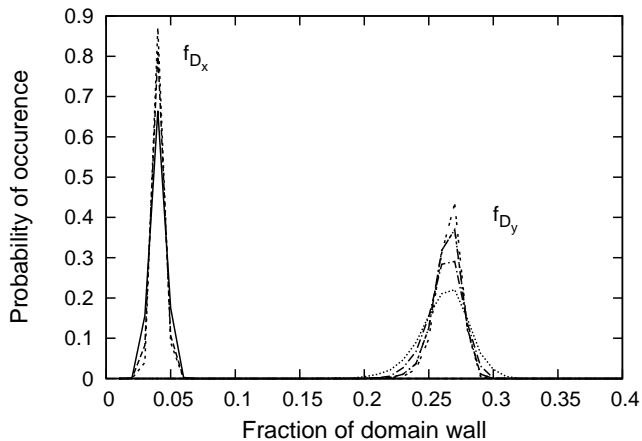}}}
\caption{
Steady state distributions of fraction  of domain walls   at $\kappa<1$ for 
different system sizes. 
The distributions become narrower as the 
system size is increased.
}
\label{distklt1}
\end{figure}

The distribution of the fraction of domain walls in the steady state shown in 
Fig. \ref{distklt1} also reveals some important features. The distribution for $f_{D_x}$ and
$f_{D_y}$ are both quite narrow with the most probable values being
$f_{D_x} \simeq 0.27$ and $f_{D_y} \simeq 0.04$  (these values are very close to the 
average values). 
With the increase in system size, the distributions tend to become 
narrower, indicating that they approach a delta function like
behaviour in the thermodynamic limit. 

\subsection{$\kappa > 1$}

It was already observed that $\kappa=1$ is the value at which the dynamical behaviour
of the ANNNI model changes drastically in one dimension. In two dimensions,
this is also true, however, we find that the additional ferromagnetic 
interaction along the vertical direction is able to affect the dynamics 
to a large extent.
Again, similar to the one dimensional case, we have different dynamical behaviour for 
$\kappa=1$ and $\kappa > 1$. In this subsection we discuss the behaviour for  $\kappa > 1$
while the $\kappa=1$ case is discussed in the next subsection.

\begin{figure}[h]
\centering
{\resizebox*{7cm}{!}{\includegraphics{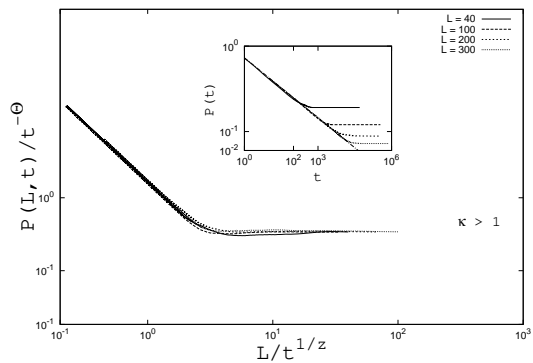}}}
\caption{The collapse of scaled persistence data versus scaled time using $\theta=0.235$ and $z=2.08$ is shown for
different system sizes for $\kappa > 1$.  Inset shows 
the unscaled data.
}
\label{collkgt1}
\end{figure}

The persistence 
probability follows a power law decay with $\theta =0.235\pm0.001$ for all $\kappa> 1$, 
while the finite size scaling analysis made according to (\ref{fss}) 
suggests a $z$ value $2.08 \pm 0.01$.
This is checked for different values of $\kappa$ ($\kappa = 1.3,1.5,2.0,20, 
100$) and 
the values of $\theta$ and $z$ have negligible variations 
with $\kappa$ which do not show any systematics.
Hence  we conclude that the exponents are independent of $\kappa$ for $\kappa > 1$.
A typical behaviour of the raw data as well as the data collapse  is shown in Fig. \ref{collkgt1}.

%

\begin{figure}[h]
\centering
\rotatebox{270}{\resizebox*{6cm}{!}{\includegraphics{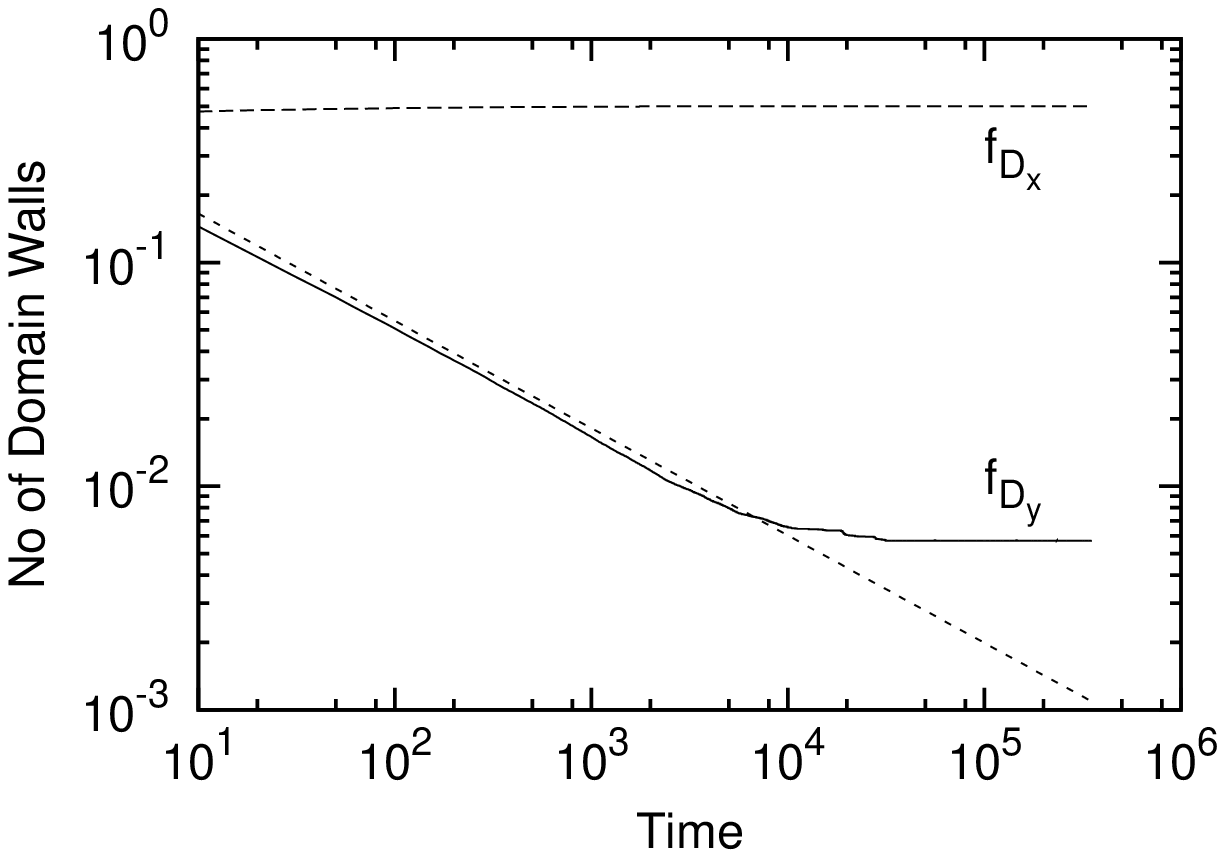}}}
\caption{
Decay of the fraction  of domain walls  with time at $\kappa>1$ are shown  along
 horizontal and vertical directions. The  dashed line has slope equal to 0.48.
}
\label{domainkgt1}
\end{figure}

The dynamics of the average fraction of domain walls along the horizontal
direction, $f_{D_x}$ again shows a fast saturation while that in the $y$ direction has a power law decay with an exponent 
$\simeq 0.48$
(Fig. \ref{domainkgt1}).
 This exponent  is also independent of $\kappa$.
As mentioned in section II, we find that there is a good agreement of the
value of this 
exponent with that of $1/z$ obtained from the finite size scaling behaviour of 
$P(t)$ implying that the average domain size $D$ is inversely proportional $f_{D_y}$. This is quite remarkable, as the fraction of domain walls 
calculated in this manner is not exactly equivalent to the inverse of
domain sizes in a two dimensional lattice; the fact that $f_{D_x}$ remains
constant may be the reason behind the good agreement (essentially the
two dimensional behaviour is getting captured along the dimension
where the number of domain walls show significant change in time).

Although the persistence and dynamic exponents are $\kappa$ independent,
we find that the distribution of the number of domain walls
has some nontrivial $\kappa$ dependence. 

\begin{figure}[h]
\centering
\rotatebox{270}{\resizebox*{5cm}{!}{\includegraphics{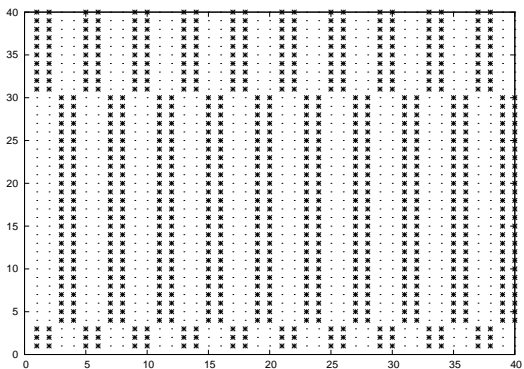}}}
\caption{A typical snapshot of a steady state configuration for $\kappa > 1$
with   flat horizontal interfaces
separating two regions of antiphase ordering (see text).
}
\label{snapkgt1}
\end{figure}

Though the system, for all $\kappa > 1$, evolves to a state with
antiphase order along the horizontal direction, the ferromagnetic order along  vertical chains 
is in some cases separated by one or more  domain walls.
A typical snapshot is shown in  Fig. \ref{snapkgt1} displaying that one essentially gets a striped state
here like in the two dimensional Ising model.

Interfaces which occur parallel to the $y$ axis, separating two
regions of antiphase and 
 keeping the ferromagnetic ordering 
along the vertical  direction intact, are extremely rare, the probability 
vanishing for larger sizes. Quantitatively this means we should get $f_{D_x}=0.5$ at long times
which is confirmed by the data (Fig. \ref{domainkgt1}). Hence in the following our discussions 
on striped state will always imply flat horizontal  interfaces, i.e.,  antiphase ordering along each horizontal 
row but the ordering can be of different types (e.g., a $++--++--\cdots$
type and a $--++--++\cdots$ type, which one can call a `shifted' 
antiphase ordering with respect to the first type). 

It is of interest to investigate whether these striped states survive in the
infinite systems. To study this, we consider the distribution of the number of
domain walls rather than the fraction for different system sizes.
The probability that there are no domain walls, 
or a perfect ferromagnetic phase along the vertical direction,
turns out to be weakly dependent on the system sizes but having
different values for different ranges of values of $\kappa$. 
For
$1< \kappa < 2$, it is $\simeq 0.632$, for $\kappa = 2.0$, it is $\simeq 0.544$
while for any higher value of $\kappa $, this probability is about 0.445. 
Thus it increases for $\kappa$ although not in a continuous manner and
like the two dimensional case, we find that there is indeed a finite
probability to get a striped state. 

\begin{figure}[h]
\centering
\rotatebox{270}{\resizebox*{6cm}{!}{\includegraphics{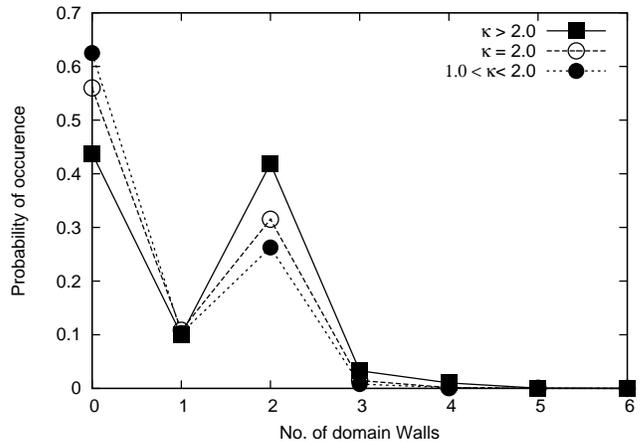}}}
\caption{
Normalised steady state distributions of number  of domain walls   for different  $\kappa>1$ show that
striped states occur with higher probability as $\kappa$ increases.
The lines are guides to the eye.
}
\label{distkgt1}
\end{figure}

While we look at the full distribution of the number of domain walls at steady state 
(Fig. \ref{distkgt1}),
 we find that there are dominant peaks at $N_{D_y}=0$ (corresponding to the unstriped state)
and at $N_{D_y}=2$ (which means there are two interfaces). 
However, we find that the distribution shows that there could be odd values 
of $N_{D_y}$ as well. This is because the antiphase has a four fold degeneracy
and the and a `shifted' ordering can occur in several ways such that 
odd values of $N_{D_y}$ are possible. In any case, the number of interfaces never exceeds 
$N_{D_y}=6$ for the system sizes considered.


\subsection{$\kappa=1$}

\begin{figure}[h]
\centering
{\resizebox*{7cm}{!}{\includegraphics{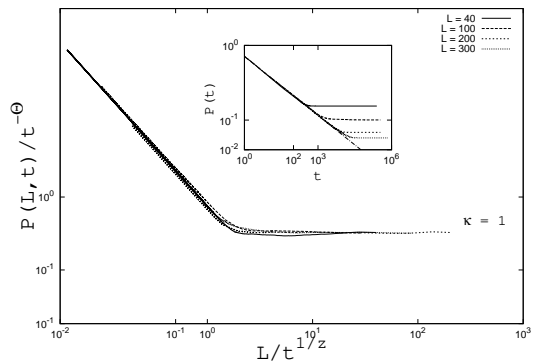}}}
\caption{The collapse of scaled persistence data versus scaled time using $\theta=0.263$ and $z=1.84$ is shown for
different system sizes at $\kappa =1$.  Inset shows
the unscaled data.
}
\label{collk1}
\end{figure}

Here  we find that the persistence 
probability follows a power law decay with $\theta =0.263\pm0.001 $. 
The finite size scaling analysis suggests a $z$ value $1.84\pm0.01$ (Fig. 
\ref{collk1}).\\

\begin{figure}[h]
\centering
\rotatebox{270}{\resizebox*{6cm}{!}{\includegraphics{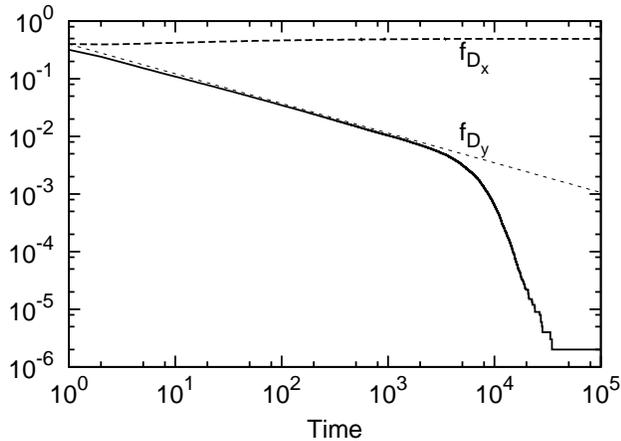}}}
\caption{Decay of the fraction  of domain walls  with time at $\kappa=1$ are shown  along
 horizontal and vertical directions. The dashed line has slope equal to 0.515.
}
\label{domaink1}
\end{figure}

We have again studied the dynamics of $f_{D_x}$ and $f_{D_y}$; the former shows a fast saturation at $0.5$ 
while the latter shows a rapid decay to  zero after an initial power law behaviour with an exponent $\approx 0.515$ (Fig. \ref{domaink1}). 
This value, unlike in the case $\kappa>1$, does not show very good agreement with  $1/z$ obtained from
the finite size scaling analysis. We will get back to this point in the next section.

The  results for $f_{D_x}$ and $f_{D_y}$ imply that the system reaches a perfect antiphase configuration 
as there are no interfaces left in the system  with $f_{D_x}=0.5$ and $f_{D_y} =0$ at
later times.

\subsection{$\kappa \leq 0.0$}

In order to make a comparison with the purely ferromagnetic case, we have also studied the 
Hamiltonian (\ref{annni2d}) with negative values of $\kappa$ which
essentially corresponds to the two dimensional Ising model with anisotropic next nearest 
neighbour 
ferromagnetic interaction.

\begin{figure}[h]
\centering
{\resizebox*{7cm}{!}{\includegraphics{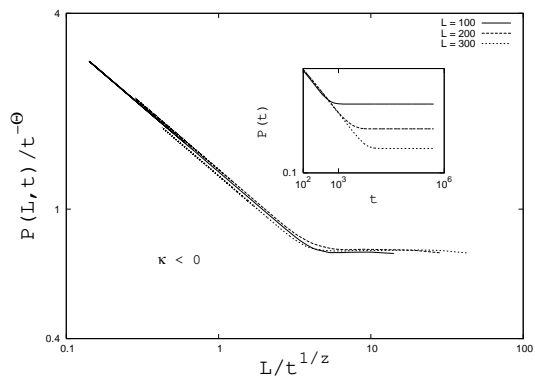}}}
\caption{The collapse of scaled persistence data versus scaled time using $\theta=0.20$ and $z=2.0$ is shown for
different system sizes for $\kappa < -1$.  Inset shows
the unscaled data.
}
\label{collklt0}
\end{figure}

$\kappa=0$ corresponds to the pure two dimensional Ising model for which
the numerically calculated value of $\theta \simeq 0.22$ is verified.
We find a new result when $\kappa$ is allowed to assume negative values, the persistence
exponent $\theta$ has a value $\simeq 0.20$ for $|\kappa| >1$ while for $0< |\kappa| \leq 1$,
the value of $\theta$ has an apparent dependence on $\kappa$, varying between 0.22 to 0.20.
However, it is difficult to numerically confirm the nature of the dependence in such a range and
we have refrained from doing it. At least for $|\kappa| >> 1$, the persistence exponent is definitely
different from that of at $\kappa = 0$.
The growth exponent $z$ however, appears to be constant and $\simeq$ 2.0 
for all values of $\kappa \leq 0$. 
A data collapse for large negative $\kappa$ is shown in Fig. \ref{collklt0} using  $\theta=0.20$ and $z=2.0$.

\begin{figure}[h]
\centering
\rotatebox{270}{\resizebox*{6cm}{!}{\includegraphics{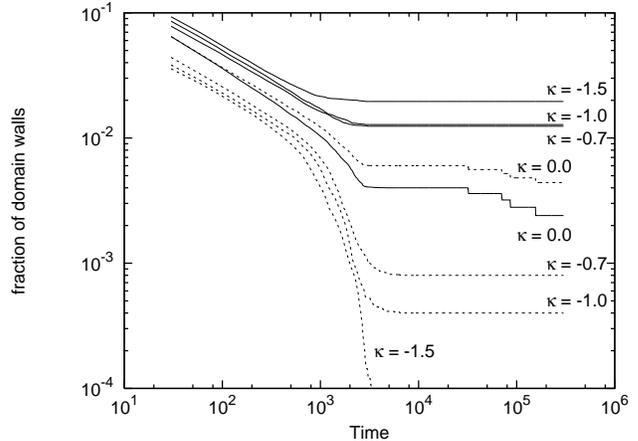}}}
\caption{
Decay of the fraction  of domain walls  with time at $\kappa \leq 0$ are shown  along
 horizontal ($f_{D_x})$, shown by dotted lines)  and vertical ($f_{D_y})$, shown by solid lines)
directions. 
}
\label{domainklt0}
\end{figure}

 The effect of the anisotropy shows up clearly 
in the behaviour of $f_{D_x}$ and $f_{D_y}$ as functions of time
(Fig. \ref{domainklt0}). For $\kappa=0$, they have identical behaviour, both
reaching a finite saturation value showing that there may be interfaces generated in either of the
directions (corresponding to the  striped states which are known to occur here). As the  absolute value of $\kappa$
is increased, $f_{D_{x}}$ shows a  fast decay to zero while  $f_{D_{y}}$ attains a constant value.
The saturation value attained by $f_{D_{y}}$ increases markedly with  $|\kappa|$ while 
 for $f_{D_{x}}$  the decay to zero becomes faster.
 One can conduct a stability analysis for striped states
to show that such states become  unstable when the interfaces are vertical
and $\kappa $ increases beyond $1$, leading to the result $f_{D_x} \to 0$.

 Extracting the $z$ value from the variations of $f_{D_{x}}$ or  $f_{D_{y}}$ is not very simple
here as the quantities do not show smooth power law behaviour over a sufficient interval of time. 

\begin{figure}[h]
\centering
\rotatebox{270}{\resizebox*{6cm}{!}{\includegraphics{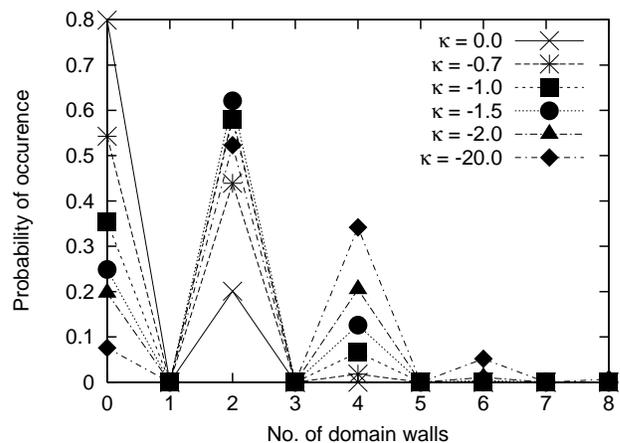}}}
\caption{
Normalised steady state distributions of number  of domain walls   for different  $\kappa \leq 0$ show that
striped states occur with higher probability as $|\kappa|$ increases.
The lines are guides to the eye.
}
\label{distklt0}
\end{figure}

The fact that $f_{D_y}$ and/or  $f_{D_x}$ reach a finite saturation value 
indicates that 
 striped states occur here as well.
The behaviour of  $f_{D_{x}}$ and  $f_{D_{y}}$ suggests that 
in contrast to the isotropic case where
interfaces can appear either horizontally or vertically, 
here
the interfaces appear dominantly along the $x$ 
direction  as $\kappa$ is increased. 
Thus the  normalised distribution of the number of domain walls along $y$ is shown in Fig. \ref{distklt0}. 
We find that as $\kappa$ is increased in magnitude, more and more interfaces
appear. However, the number of interfaces is always even consistent with the
fact that interfaces occur between ferromagnetic domains of all up and all down spins.

\begin{figure}[h]
\centering
\rotatebox{270}{\resizebox*{6cm}{!}{\includegraphics{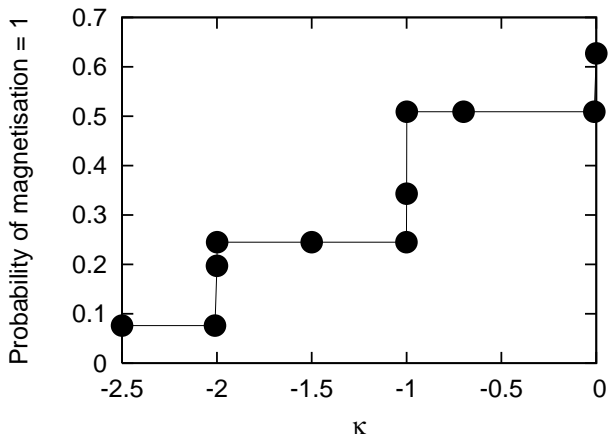}}}
\caption{
Probability that the magnetisation takes a steady state value equal to unity is shown against
$\kappa$ when $\kappa \leq 0$.
}
\label{distmag}
\end{figure}

Lastly in this section, we discuss the behaviour of the magnetisation
which is the order parameter in a ferromagnetic system. As striped states are formed,
the magnetisation will assume values less than unity.  The probability of 
configurations with magnetisation equal to unity shows a stepped behaviour, 
with values changing at $|\kappa|=1$ and $2$ and assuming constant values at
 $1< |\kappa|<2$ and above $|\kappa| = 2$ (Fig. \ref{distmag}).

\section{Discussions and Conclusions}

We have investigated some dynamical features of the ANNNI model in two dimensions 
following a quench to zero temperature. 
We have obtained the results that the dynamics is very much dependent on the 
value of $\kappa$, the ratio  of the antiferromagnetic interaction to the 
ferromagnetic interaction along one direction. This is similar to the 
dynamics of the one dimensional model studied earlier, but here we have more 
intricate features, e.g., that of the occurrence of quasi frozen-in structures
for $\kappa < 1$ where the persistence probability shows a very slow decay
with time. 
Persistence probability is
algebraic for $\kappa\geq 1$, but exactly at $\kappa=1$, the 
exponents $\theta$  and $z$ are different from those at $\kappa > 1$. The exponents for $\kappa > 1$ 
are in fact
 very close to those of the two dimensional
Ising model with nearest neighbour ferromagnetic interaction.
(This was not at all true for the one dimensional ANNNI chain, where the
persistence exponent at $\kappa > 1$ was found to be appreciably different from
that of the one dimensional Ising chain with nearest 
neighbour ferromagnetic interaction.) 
This shows that the ferromagnetic interaction along the vertical direction
is able to negate the effect of the antiferromagnetic interaction 
to a great extent. This is apparently  a  counter intuitive phenomenon,
$\kappa =0$ and $\kappa>1$ having very similar dynamic behaviour while
in the intermediate values, the dynamics is qualitatively and quantitatively
different. At far as dynamics is concerned, the ANNNI model
in two dimensions cannot be therefore treated perturbatively. 

Although the values of $\theta$ and $z$ are individually quite close  
for $\kappa=0$ and $\kappa> 1$, the product $z\theta= \alpha $ are quite different. 
For $\kappa=0$, $\alpha \simeq 0.44$ while for 
$\kappa > 1$, it is  $0.486 \pm 0.002$. This shows that the spatial
correlations of the persistent spins are quite 
different for the two and one can safely say that the dynamical
class for $\kappa=0$ and $\kappa > 1$ are not the same.
$\kappa=1$ is the special point where the
dynamic behaviour changes radically.  
Here there appears to be some ambiguity regarding the value of $z$; estimating $\alpha$ from the finite size 
scaling analysis gives $\alpha \approx 0.484 \pm 0.005$ while using the $z$ value from the domain dynamics, the estimate is
approximately equal to 0.51. However,  the dynamics of the domain sizes
may not be very accurately reflected by the dynamics of $f_{D_y}$ in which case $\alpha \approx 0.48$
is a more reliable result. Thus we find that although the values of $\theta$ and $z$ are quite different
for $\kappa =1$ and $\kappa >1$, the $\alpha$ values are  close.

We would  like to add here  that when there is a power law
decay of a quantity related to the domain dynamics, it is highly unlikely that it will be accompanied by 
an exponent which is different from the growth exponent. Thus, even though we get
slightly different values of $z$ for $\kappa =1$ from the two analyses, it is 
 more likely that this is an artifact of the
numerical simulations.

Another feature present in the 
two
 dimensional Ising model is the finite probability with which it ends up in a striped state.
The same happens for $\kappa>1$, but here the probabilities are quite different 
and also dependent on $\kappa$. We find that there is a significant role 
of the point $\kappa=2$ here as this probability has different values 
at  $\kappa =2$,  $\kappa > 2$ and $\kappa < 2$.

Comparison of the ANNNI dynamics with that of the ferromagnetic anisotropic Ising model shows some interesting features.
In the latter, one gets a new  value of persistence exponent for $\kappa < -1$ while in the former a new value is obtained 
for  $\kappa \geq 1$.The new values    (except for $\kappa=1$) 
are in fact very close to that of the two dimensional Ising model, but simulations
done for identical system sizes averaged over the same number of initial configurations are able
to confirm the difference. 
  The 
qualitative behaviour of the domain dynamics is again strongly $\kappa$ dependent when
$\kappa$ is negative.
Another point to note 
is that  the probability that the system evolves to a pure state is $\kappa$ dependent  in both the ANNNI
model and the Ising  model. 
In both cases in fact, this probability decreases in a step like manner with increasing magnitude of $\kappa$.
We also find the interesting result that while the distribution of the number of domain walls
can have non-zero values at odd values of $N_D$  in the ANNNI model because of the four fold degeneracy of the antiphase,
for the Ising model,   odd values of $N_D$ are
not permissible as the ferromagnetic phase is two fold degenerate.

Finally we comment on the fact that although the dynamical behaviour, as far as domains  are 
concerned, reflects the inherent anisotropy of the system (in both the
ferromagnetic and antiferromagnetic models), the persistence probability is unaffected by it. In order to verify 
this, we estimated $P(t)$ along an isolated   chain of spins along $x$ and $y$ directions separately and 
found that the two estimates gave identical results  for all values of $\kappa$.

In conclusion, it is found that except for the region $0 < |\kappa| <1$, the dynamical
behaviour of the Hamiltonian (\ref{annni2d}) is remarkably similar for negative and positive $\kappa$; 
the persistence and growth exponents get only marginally affected compared to the values   of
 the two
dimensional Ising case ($\kappa=0$) and the domain distributions have similar nature.
However, the region $0 < \kappa < 1$ is extraordinary, where algebraic decay of persistence is absent. There is
dynamic frustration
as the system gets locked in a metastable state consisting of ladder-like domains and the dynamics
is very slow because of the presence of  quasi-frozen structures. 
There is in fact dynamic frustration at other $\kappa$ values also in the sense that except for $\kappa=1$,
the system has a tendency to get locked in a ``striped state''. However, even in that case, the algebraic
decay of the persistence probability is observed.  Thus algebraic decay of 
persistence probability seems to be valid only when the metastable state is a striped state. 
Although there is no dynamic frustration at $\kappa=1$ in the sense that it always evolves to a state with perfect
antiphase structure,
it happens to be a very special point where the persistence exponent and growth exponents are
unique and appreciably different from those of the $\kappa=0$ case. 

In this paper, the behaviour of the two dimensional ANNNI model under a zero temperature has been discussed; the dynamics
at finite temperature can be in fact quite different. At finite temperatures,  the spin flipping probabilities are stochastic,  
and the  dynamical frustration may be overcome by the
thermal fluctuautions. It has been observed earlier \cite{pratap}  that in a  thermal annealing scheme of the one dimensional
 ANNNI model,  the $\kappa=0.5$ point becomes significant.  A similar effect can occur for the two dimensional case as
well. The definition of persistence being quite different at finite temperatures \cite{derrida2}, it is also not easy to
guess its behaviour (for either the one or two dimensional model) simply from the results of the  zero temperature quench. Indeed,  the ANNNI model under a finite temperature 
quench  is an open  problem which could  be  addressed in the future.  

\medskip

Acknowledgments:
The authors thank Purusattam Ray for discussions.
SB  acknowledges financial support from UGC grant no. UGC/520/JRF(RFSMS)  and computational facility 
from DST FIST. 
The work of AKC was supported by the Centre for Applied 
Mathematics and Computational Science (CAMCS) of the Saha Institute of
Nuclear Physics. 
PS acknowledges financial support from CSIR grant no. 3(1029)/05-EMR-II.

\end{document}